\documentclass[aps,prl,preprint,groupedaddress]{revtex4-2}
\usepackage{mathtools,amsmath,graphicx,array,enumitem,xcolor,subcaption,hyperref,tabularx,ragged2e,amssymb,float,booktabs}
\usepackage{natbib}
\begin{document}

\title{Constraints on Lorentz Invariance Violation from GRB 221009A Using the DisCan Method}

\author{Yu Xi$^{1,2}$}
\author{Fu-Wen Shu$^{1,2,3}$}
\thanks{shufuwen@ncu.edu.cn}  
\affiliation{
$^{1}$Department of Physics, Nanchang University, Nanchang, 330031, China\\
$^{2}$Center for Relativistic Astrophysics and High Energy Physics, Nanchang University, Nanchang, 330031, China\\
$^{3}$Center for Gravitation and Cosmology, Yangzhou University, Yangzhou, 225009, China
}

\begin{abstract}
Lorentz symmetry is a cornerstone of modern physics, and testing its validity remains a critical endeavor. In this work, we analyze the photon time-of-flight and time-shift data from LHAASO observations of Gamma-Ray Burst GRB 221009A to search for signatures of Lorentz violation. We employed the DisCan (dispersion cancellation) method with various information entropies as cost functions, designating the results obtained with Shannon entropy as our representative outcome. This choice is attributed to the parameter-free statistical properties of Shannon entropy, which has demonstrated remarkable stability as we continually refine and enhance our methodology. In the absence of more detailed data and physical context,  it provides more stable and reliable results. We constrain the energy scale associated with Lorentz invariance violation. Our results yield 95\% confidence level lower limits of $E_{\text{QG},1} > 5.4 \times 10^{19} \, \text{GeV}$ (subluminal) and $E_{\text{QG},1} > 2.7 \times 10^{19} \, \text{GeV}$ (superluminal) for the linear case ($n$=1), and $E_{\text{QG},2} > 10.0 \times 10^{12} \, \text{GeV}$ (subluminal) and $E_{\text{QG},2} > 2.4 \times 10^{12} \, \text{GeV}$ (superluminal) for the quadratic case ($n$=2). Subsequently, we incorporated 
WCDA photons and the Knuth binning method to further optimize and complement our approach, while also performing filter using information entropies. Furthermore, we demonstrate that employing different information entropy measures as cost functions does not alter the order of magnitude of these constraints. 
\end{abstract}

\maketitle
\newpage
\section{Introduction}
Lorentz symmetry represents a fundamental principle in modern physics, serving as a cornerstone of both special relativity and quantum field theory.  However, despite the high-precision testing of symmetry across various contexts, there exists significant motivation to explore potential violations. A key motivation is the quest to resolve a fundamental challenge in modern physics: the reconciliation of General Relativity (GR) with Quantum Field Theory (QFT). While GR and QFT have achieved notable empirical success, they leave crucial questions unanswered, such as the emergence of spacetime singularities in GR. Addressing these issues necessitates a comprehensive quantum theory of gravity. Several Quantum Gravity (QG) models suggest the potential violation of Lorentz symmetry at scales where QG effects become significant, as evidenced by various studies \cite{Amelino-Camelia:1997ieq,Amelino-Camelia:1996bln,Ellis:2003sd,Gambini:1998it,Carroll:2001ws,PhysRevD.61.027503,colladay1998lorentz,amelino2000quantum}. This provides a compelling rationale for investigating Lorentz Invariance Violation (LIV) (for an overview, refer to \cite{Liberati:2009pf}).

Recent advancements in high-energy astrophysical observations provide extra motivations of testing LIV beyond theoretical conjectures. Notably, the subtle spectral deviations between the Greisen-Zatsepin-Kuzmin (GZK) suppression features and theoretical predictions \cite{abraham2008observation,abu2013cosmic,abbasi2023energy}, alongside the anomalous attenuation deficit of TeV gamma rays from Active Galactic Nuclei (AGN), termed the TeV gamma-ray crisis \cite{Protheroe:2000hp}, serve as pivotal observational probes due to their acute sensitivity to potential LIV \cite{Toma:2012xa}. These astrophysical phenomena serve as pivotal avenues for detecting potential violations of Lorentz symmetry. They offer essential insights into exploring deviations from standard dispersion relations in high-energy contexts.  Pioneering work by researchers like collaborations such as MAGIC \cite{albert2008probing,acciari2020bounds,ahnen2017constraining,albert2007variable,Zhang:2020bzg}  have established increasingly stringent limits on potential Lorentz symmetry violations, particularly in high-energy astrophysical contexts. Our work is inspired by the MAGIC collaboration's efforts \cite{albert2008probing}, aiming to verify whether similar constraints can be obtained in different source scenarios.

The DisCan (Dispersion Cancellation) method \cite{scargle2008algorithm} emerges as a powerful technique for investigating LIV, does not require a pre-specified light curve model, in contrast to the maximum likelihood method. Utilizing the comprehensive photon dataset from the Large High Altitude Air Shower Observatory (LHAASO) \cite{lhaaso2023tera}, specifically the KM2A (Kilometer Square Array) and WCDA (Water Cherenkov Detector Array) detector. KM2A is a ground-based particle detector array covering approximately one square kilometer, composed of electromagnetic particle detectors and muon detectors. Its primary strength lies in high-sensitivity detection of gamma rays and cosmic rays in the energy range from tens of TeV to PeV (peta-electronvolts), enabling effective discrimination between cosmic-ray backgrounds and gamma-ray signals. WCDA comprises three large water pools with a total area of 78,000 square meters, utilizing Cherenkov light radiation to cover the 100 GeV to tens of TeV energy range. Key advantages of WCDA include its wide field of view, nanosecond-level timing resolution, and all-weather monitoring capability, making it particularly adept at capturing rapid variability in high-energy astrophysical phenomena. This method offers a novel approach to constraining potential Lorentz symmetry breaking. The motivation stems from the need to probe fundamental physics at energy scales approaching quantum gravity thresholds.

In this paper we focuse on analyzing the time-of-flight measurements of photons from GRB 221009A \cite{yang2024constraints}, one of the brightest gamma-ray bursts ever detected. By applying the DisCan method in conjunction with various information entropies as cost functions, and utilizing the Knuth binning method \cite{knuth2019optimal} as an optimization of the original fixed equal-bin distribution, aiming to provide improved constraints on potential LIV.  Through the comparison of different information entropies, such as Shannon entropy, Rényi entropy, and Tsallis entropy, we derive effective limits on LIV that are consistent with previous NASA findings \cite{scargle2008algorithm}, indicating that Shannon entropy as a cost function yields higher credibility. More significantly, we extrapolate this approach to long-duration bursts and high-energy regimes, not just with simulated data, thus expanding the application domain of the DisCan method. 

The structure of this paper is as follows : Section II presents the theoretical background and methodological approach. Section III details the DisCan method and its implementation. Section IV discusses the results and their implications. Finally, Section V offers our  discussions conclusions 

\section{Methods and Approaches}
Quantum gravity theories propose that space-time exhibits quantum fluctuations, which may lead to a non-trivial refractive index that influences the propagation of particles \cite{ellis2008derivation}. Different quantum gravity models suggest that the speed of particle propagation can vary with energy, a phenomenon referred to as LIV.

A fundamental expectation associated with quantum gravity is a modified photon dispersion relation \cite{piran2024lorentz} :

\begin{equation}
E^2 = p^2c^2 \left[1 + \sigma \left(\frac{E}{\varepsilon_{QG,n}^{(\sigma)}E_{Pl}}\right)^n\right],
\end{equation}
where $E_{Pl} \approx 1.22 \times 10^{19}$ GeV denotes the Planck energy. The parameter $\varepsilon_{QG,n}^{(\sigma)}$ quantifies the LIV energy scale relative to $E_{Pl}$. Typically, the value of $n$ takes 1 or 2, corresponding to first-order or second-order breaking energy scales, respectively. The index $\sigma = \pm1$ indicates the direction of dispersion modification.

This dispersion relation introduces an energy-dependent photon velocity. In the regime where $E \ll \varepsilon_{QG,n}E_{Pl}$, we can express the relationship as :

\begin{equation}
v = c \left[ 1 + \sigma \frac{n+1}{2} \left( \frac{E}{\varepsilon_{QG,n}^{(\sigma)} E_{Pl}} \right)^n \right].
\end{equation}
Photon propagation can exhibit either subluminal or superluminal characteristics, which correspond to $\sigma = -1$ and $\sigma = +1$, respectively. This energy-dependent velocity results in a time of flight (TOF) for a photon with energy $E$ that differs from what would be expected if it traveled at the conventional speed of light \cite{jacob2008lorentz}.

The modified TOF can be mathematically represented as \cite{jacob2008lorentz} :

\begin{equation}
\label{eq:t}
\delta t_{LIV}(E) = -\sigma \frac{n+1}{2H_0} \left(\frac{E}{\varepsilon_{QG,n}^{(\sigma)} E_{Pl}}\right)^n 
\times \int_0^z \frac{(1 + \zeta)^n d\zeta}{\sqrt{\Omega_M(1 + \zeta)^3 + \Omega_\Lambda}},
\end{equation}
where $z$ denotes the redshift of the source. The Hubble-Lemaître constant, denoted as $H_0$, is taken to be approximately 67.5 km s$^{-1}$ Mpc$^{-1}$. The current fraction of matter in the Universe is represented by $\Omega_M = 0.315$, while the cosmological constant fraction is $\Omega_\Lambda = 0.685$ \cite{aghanim2020planck}. Specifically, for the case of GRB 221009A, which has a redshift of $z = 0.151$ \cite{lhaaso2023very}, the relevant values are utilized in the expression.

In this work, which also makes use of the publicly accessible data from LHAASO for an initial exploration, we took a different approach from Piran's phenomenological average correction constant. Specifically, we utilized the complete time-energy information data of photons detected by KM2A \cite{lhaaso2023very} during GRB 221009A and applied a DisCan method to impose observational constraints on the potential Lorentz violation energy scale. The results obtained from our analysis mutually validate the findings of Piran's work.

The DisCan method is an algorithm for detecting quantum gravity dispersion in photons. The algorithm adjusts the arrival times of photons to cancel out the dispersion caused by quantum gravity effects, thereby detecting the energy-dependent time delays. Assuming the energy-dependent time delay is given by \cite{scargle2008algorithm} :
\begin{equation}
\label{eq:lin}
t_{\text{obs}} = t_{\text{true}} + \theta_1\cdot E,
\end{equation}
where $t_{\text{obs}}$ is the observed arrival time, and $t_{\text{true}}$ denotes the time at which the photon would have arrived had it suffered no quantum gravity time shift \cite{scargle2008algorithm}. $\theta_1$ is the dispersion coefficient when $n=1$ and $E$ is the photon energy, with the time unit in $s$ and the dispersion coefficient unit in $s$/TeV. Similarly, we can also assume a quadratic energy-dependent time delay as :
\begin{equation}
\label{eq:quad}
t_{\text{obs}} = t_{\text{true}} + \theta_2\cdot E^2,
\end{equation}
where $\theta_2$ is the dispersion coefficient when $n$=2, with the dispersion coefficient unit in $s$/TeV$^2$.

For both the first-order and second-order delay cases, we construct a time profile using the adjusted arrival times. Then, we optimize the sharpness of the time profile to find the best dispersion coefficient $\theta_i$, using Shannon entropy, Rényi entropy and tsallis entropy.

Based on the previous formulas \eqref{eq:t} \eqref{eq:lin} \eqref{eq:quad} and numerical values, and $\delta t_{LIV}(E) = t_{\text{obs}} - t_{\text{true}},$ we can derive :
\begin{align}
\varepsilon_{QG,1}^{(\sigma)} &\simeq -5.7\sigma / \theta_1 & n &= 1, \\
\varepsilon_{QG,2}^{(\sigma)} &\simeq -7.5\sigma / (10^8 \cdot \theta_2) & n &= 2
\end{align}
Since $\varepsilon_{QG}$ is always positive, we can determine $\sigma$ through the parameter $\theta_i$, thereby determining whether it is superluminal or subluminal.

The advantage of the DisCan method is its applicability to arbitrary dispersion models and its high sensitivity. By adjusting the photon arrival times, the DisCan method can effectively detect the small time delays caused by quantum gravity effects.
We investigate time-of-flight constraints on LIV derived from KM2A’s observations of TeV photons from GRB 221009A, the brightest gamma-ray burst, and appropriately incorporated the low-energy photon data observed by WCDA.

\section{Analysis of DisCan}
The DisCan method is applicable to any dispersion model, as well as capable of detecting small time delays caused by quantum gravity effects. For this paper, after constructing the linear dispersion model, we first choose a cost function for the overview of time after binning, in order to optimize the selection of parameter $\theta_1$ , $\theta_2$ and obtain the corresponding LIV parameters $\varepsilon_{QG,1}^{(\sigma)}$ , $\varepsilon_{QG,2}^{(\sigma)}$. Finally, the breaking energy scales for $E_{\text{QG},1}$ and $E_{\text{QG},2}$ are obtained.

The LIV parameters $\theta_i$ ($i=1,2$) are estimated by maximizing the information content of the reconstructed time profiles using transformed photon arrival times through dispersion relations \eqref{eq:lin} and \eqref{eq:quad}. We employ three information-theoretic measures as cost functions :

\begin{itemize}
    \item \textbf{Shannon entropy} 
    
    The Shannon entropy is defined \cite{shannon1948} : 
    \begin{equation}
        H_{\mathrm{S}} = - \sum_{n} p_n \log p_n
        \label{eq:shannon}
    \end{equation}
    where $p_n \equiv x_n/\sum x_n$ represents the normalized photon count probability \cite{scargle2008algorithm}, with $x_n$ denoting the photon intensity in temporal bins. As the foundational measure of information uncertainty, it achieves maximal sensitivity for peaked distributions. Subsequent entropys also continue to employ these two concepts.
    
    \item \textbf{Rényi entropy} 
    
    The Rényi entropy is given \cite{renyi1961measures} :
    \begin{equation}
        H_{\mathrm{R}}(\alpha) = \frac{1}{1-\alpha} \log\left( \sum_{n} p_n^\alpha \right)
        \label{eq:renyi}
    \end{equation}
    This generalized entropy reduces to Shannon entropy when $\alpha \to 1$, with $\alpha > 1$ emphasizing high-probability bins and $\alpha < 1$ enhancing sensitivity to rare events.
    
    \item \textbf{Tsallis entropy} 
    
    The expression for Tsallis entropy is as follows \cite{tsallis1988possible} :
    \begin{equation}
        H_{\mathrm{T}}(q) = \frac{1}{q - 1} \left( 1 - \sum_{n} p_n^q \right)
        \label{eq:tsallis}
    \end{equation}
    It is characterized by non-extensive parameter $q \neq 1$, and has the same regulatory effect as $\alpha$. Originally developed for non-equilibrium thermodynamics, this measure demonstrates superior performance in systems with long-range correlations.
\end{itemize}

When entropy is minimized, the information at the source is most certain, and it is most likely to obtain the true time profile, thereby obtaining the broken parameter.

\section{result}
We utilize the publicly available KM2A observational data from the LHAASO collaboration \cite{lhaaso2023very} to investigate the values of the parameters $\theta_i$ obtained under first and second order linear models, thereby deriving the corresponding energy scale $E_{\text{QG},n}$ for LIV. Initially, we utilized only the photon time-energy data from KM2A and referenced the 10-second binning provided by LHAASO to maintain the original equal-bin distribution \cite{lhaaso2023very} . Subsequently, we incorporated additional photon data from WCDA and replaced the original equal-bin distribution with the Knuth binning method. The Knuth binning method is a type of equal-bin method that employs a piecewise constant density model with equal bin widths. It utilizes the maximum Bayesian posterior probability to directly quantify the likelihood of binning. Its primary advantage is its adaptability to any distribution, including multi-peaked, asymmetric, and complex structures, enabling it to effectively capture complex temporal structures. Notably, this method requires little to no prior information \cite{knuth2019optimal}. Our approach to handling WCDA photon data also referenced the methodology of the LHAASO collaboration \cite{yang2024constraints} .

Since WCDA photons are initially assumed to possess identical energy, with any subsequent variations in energy attributable solely to statistical fluctuations, it is posited that these photons do not enhance the clarity of the source reconstruction. Instead, they substantially diminish the influence of KM2A photons. Consequently, it is deemed appropriate to subtract the information entropy generated by WCDA photons following the binning process. This leads to a natural adoption of a refined approach, wherein the information entropy, originally defined by the combined contributions of KM2A and WCDA photons, is adjusted by subtracting the information entropy derived exclusively from WCDA photons. This adjustment effectively replaces the traditional notion of information entropy with the concept of relative information entropy. 

To account for photon energy uncertainties, we apply a Monte Carlo-based blurring technique to the photon energies, assuming Gaussian errors. In our Monte Carlo simulations ($n_{\mathrm{sim}} = 1000$), we model energy measurement uncertainties using asymmetric Gaussian distributions. For each data point $i$, positive errors follow $\mathcal{N}(\mu = 0, \sigma = \sigma_{+i})$ and negative errors follow $\mathcal{N}(\mu = 0, \sigma = \sigma_{-i})$, where $\sigma_{+i}$ and $\sigma_{-i}$ are the observed upper and lower $1\sigma$ error bounds respectively. For each realization, we generate random variates $z \sim \mathcal{N}(0,1)$ and compute the error term as:
\begin{equation*}
    \delta_i = 
    \begin{cases} 
        \sigma_{+i} \cdot z & z \geq 0 \\
        \sigma_{-i} \cdot |z| & z < 0 
    \end{cases}
\end{equation*}
Synthetic energy samples are then generated through the transformation:
\begin{equation*}
    E_{\mathrm{sample},i} = \max(0, E_{\mathrm{obs},i} + \delta_i).
\end{equation*}
We construct the histogram distribution of the parameter $\theta_i$. The 95\% confidence interval is indicated by a dashed line in the plot.

\begin{figure}[!htbp]
\centering

\begin{subfigure}[b]{0.48\textwidth}
\centering
\includegraphics[width=0.95\linewidth]{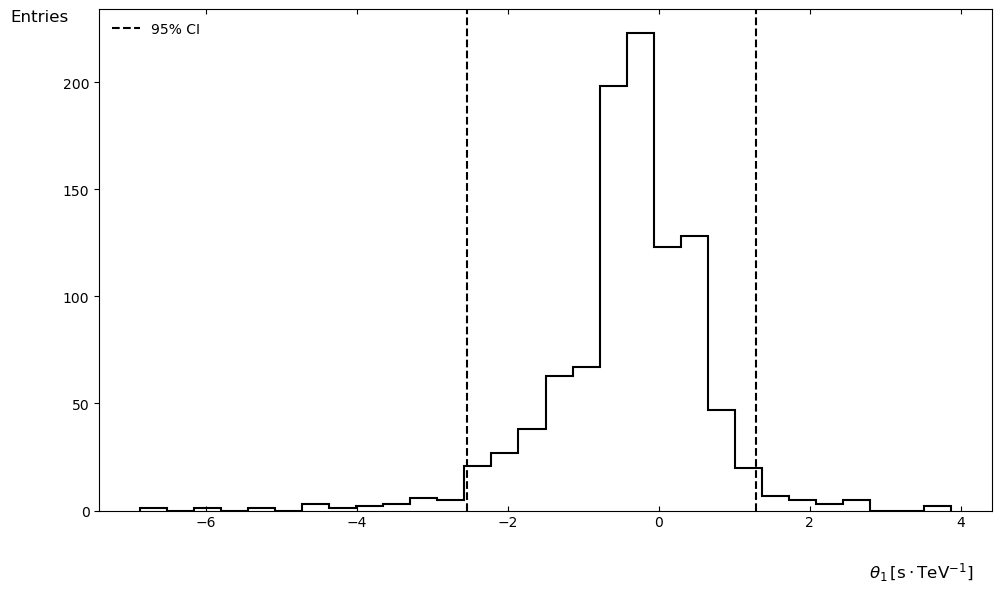}
\caption{Shannon entropy ($\alpha \to 1$)}
\label{fig:shannon_theta1}
\end{subfigure}
\hfill
\begin{subfigure}[b]{0.48\textwidth}
\centering
\includegraphics[width=0.95\linewidth]{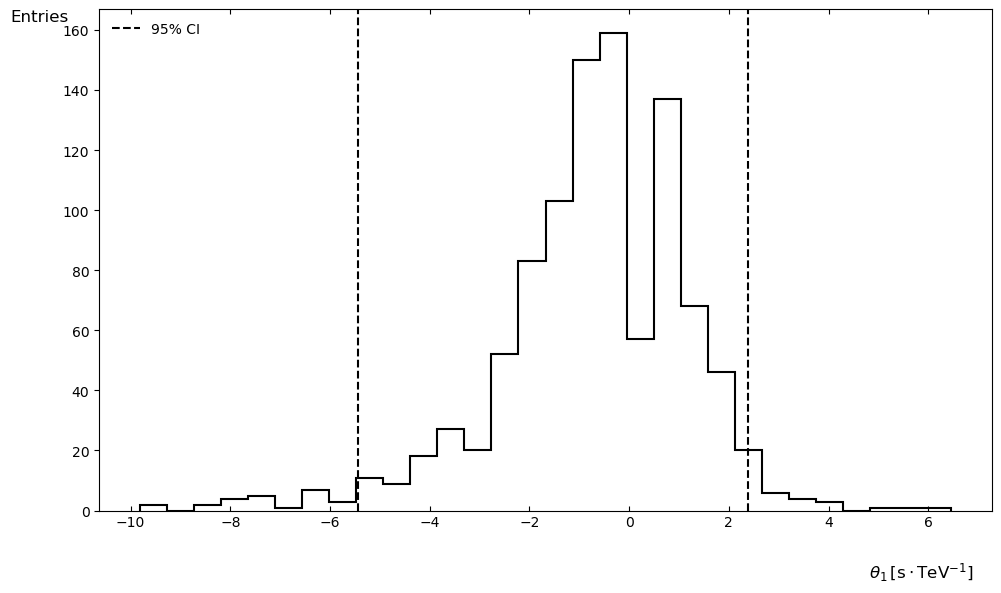}
\caption{Rényi entropy ($\alpha=2$)}
\label{fig:renyi_theta1}
\end{subfigure}

\begin{subfigure}[b]{\textwidth}
\centering
\includegraphics[width=0.6\linewidth]{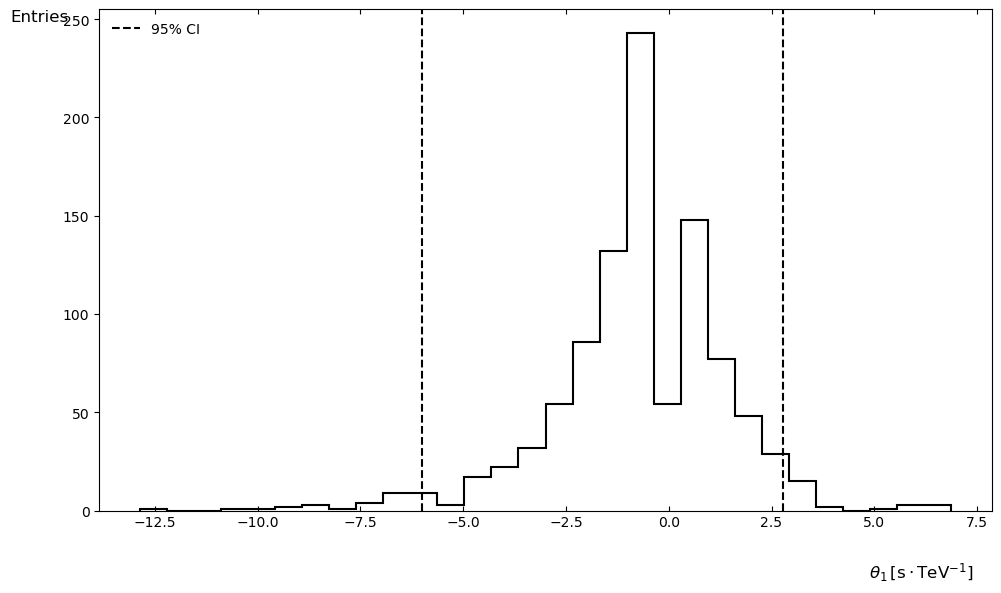}
\caption{Tsallis entropy ($q=2$)}
\label{fig:tsallis_theta1}
\end{subfigure}

\captionsetup{
    justification=RaggedRight}

\caption[LIV parameter $\theta_1$ distributions with different entropy measures]{
Probability density distributions of the first-order Lorentz invariance violation parameter $\theta_1$ estimated through: 
(a) Shannon entropy (baseline case), 
(b) Rényi entropy with $\alpha=2$, and 
(c) Tsallis entropy with $q=2$. 
Vertical dashed lines indicate 95\% confidence intervals. All histograms contain 1,000 Monte Carlo realizations with identical color mapping and axis scaling.
}
\label{fig:liv_theta1_distributions}
\end{figure}

\begin{figure}[!htbp]
\centering

\begin{subfigure}[b]{0.48\textwidth}
\centering
\includegraphics[width=0.95\linewidth]{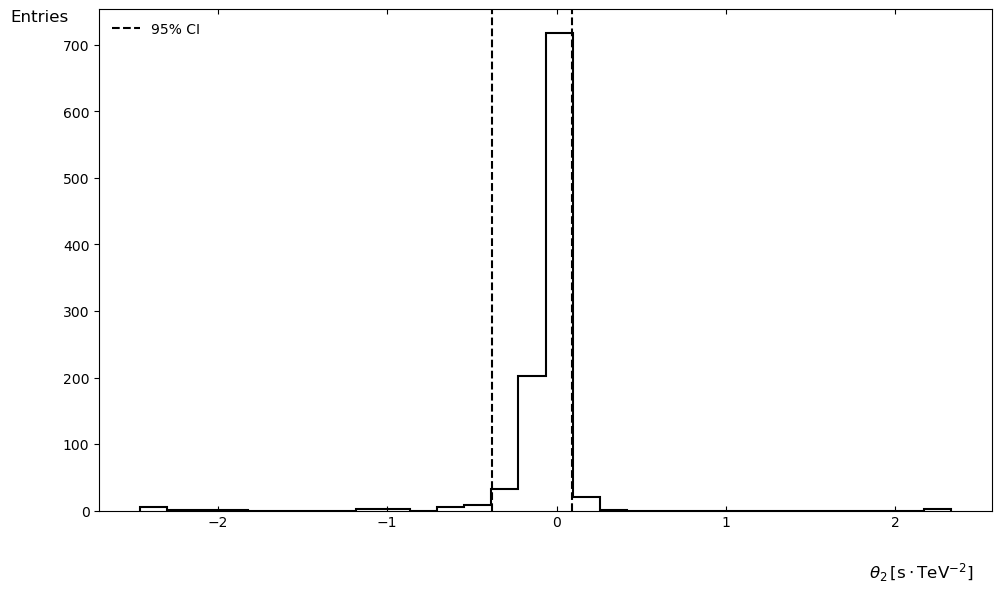}
\caption{Shannon entropy ($\alpha \to 1$)}
\label{fig:shannon_theta2}
\end{subfigure}
\hfill
\begin{subfigure}[b]{0.48\textwidth}
\centering
\includegraphics[width=0.95\linewidth]{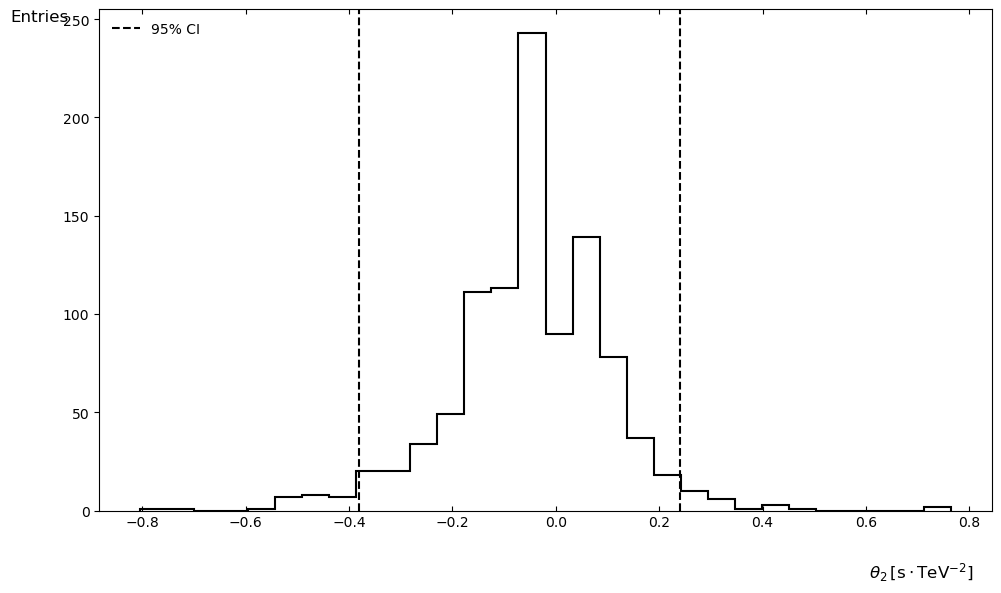}
\caption{Rényi entropy ($\alpha=2$)}
\label{fig:renyi_theta2}
\end{subfigure}

\begin{subfigure}[b]{\textwidth}
\centering
\includegraphics[width=0.6\linewidth]{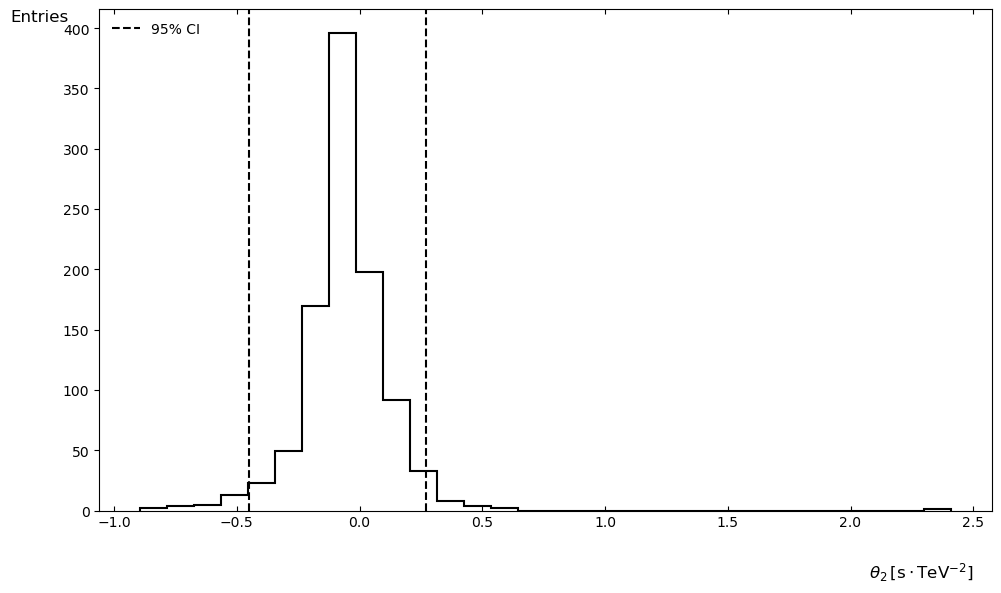}
\caption{Tsallis entropy ($q=2$)}
\label{fig:tsallis_theta2}
\end{subfigure}

\captionsetup{
    justification=RaggedRight}

\caption[LIV parameter $\theta_2$ distributions with different entropy measures]{
Probability density distributions of the second-order Lorentz invariance violation parameter $\theta_2$ estimated through: 
(a) Shannon entropy (baseline case), 
(b) Rényi entropy with $\alpha=2$, and 
(c) Tsallis entropy with $q=2$. 
Vertical dashed lines indicate 95\% confidence intervals. All histograms contain 1,000 Monte Carlo realizations with identical color mapping and axis scaling.
}
\label{fig:liv_theta2_distributions}
\end{figure}

FIGs. \ref{fig:liv_theta1_distributions} and \ref{fig:liv_theta2_distributions}  show the histogram distributions of the parameters $ \theta_1 $ and $ \theta_2 $ obtained from 1000 Monte Carlo samples, with dashed lines indicating the 95\% confidence intervals. It can be observed that, when using Shannon entropy as the cost function, the resulting parameters $ \theta_i $ exhibit shorter confidence intervals and higher concentration in the same order, without the introduction of additional parameters. Therefore, we recommend the results obtained from Shannon entropy as the summary. Nevertheless, the results obtained using Rényi and Tsallis entropy remain significant. They offer insights that complement findings from other entropy measures, such as Shannon entropy. Also, as both our linear energy-dependent time delay formulas Eq. \eqref{eq:lin} and \eqref{eq:quad} and the use of information entropy as a cost function ignore some deeper physical considerations, such as the omission of extragalactic background light (EBL) absorption in the propagation model, source-intrinsic energy-dependent photon emission delays, and other unmodeled systematic uncertainties. Nevertheless, they still hold significance as observable tests for quantum gravity theories and comparative studies with other Lorentz symmetry violations. 

\begin{figure}[htbp]
  \centering
  \begin{subfigure}{0.45\textwidth}
    \includegraphics[width=\linewidth]{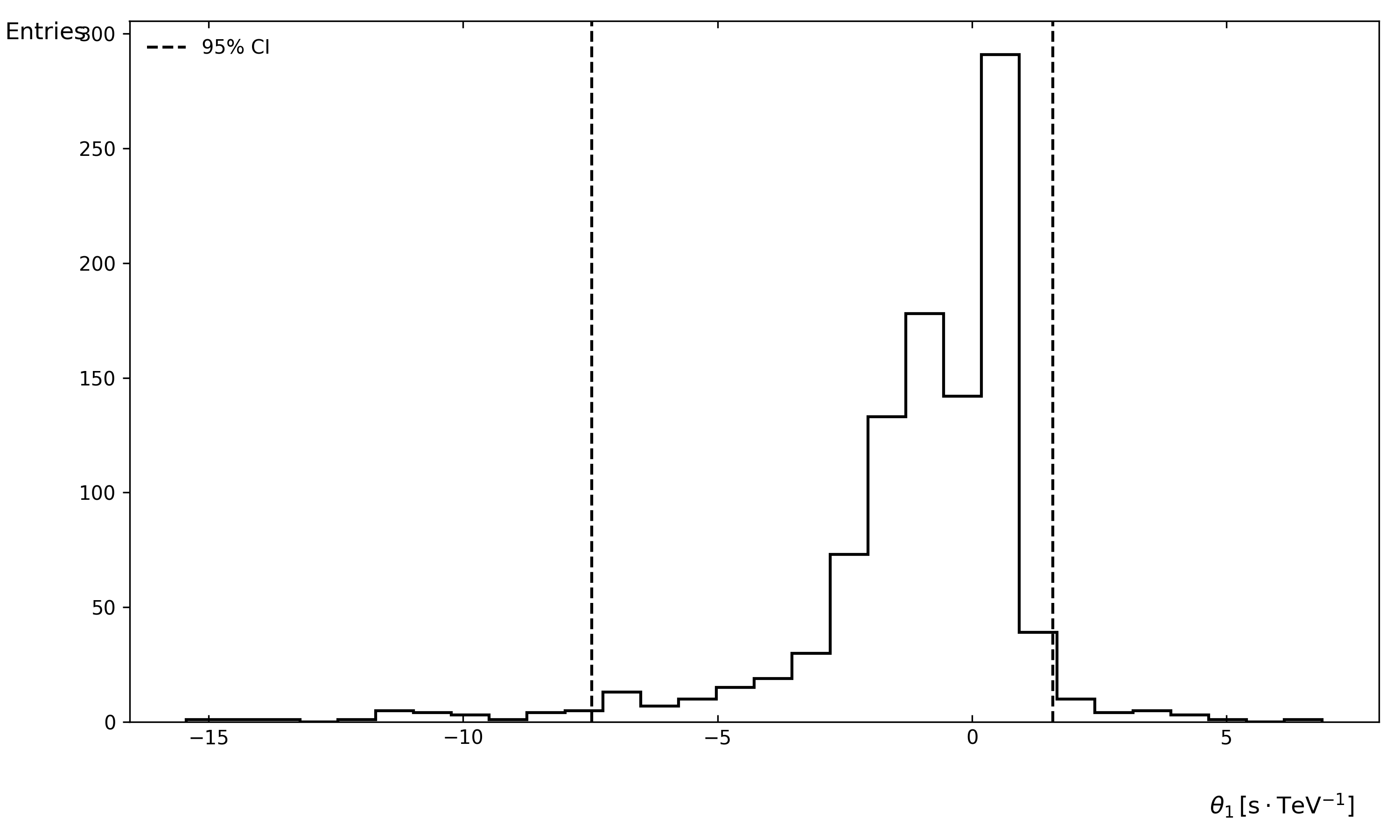}
    \caption{}
    \label{fig:bgS1}
  \end{subfigure}
  \hfill
  \begin{subfigure}{0.45\textwidth}
    \includegraphics[width=\linewidth]{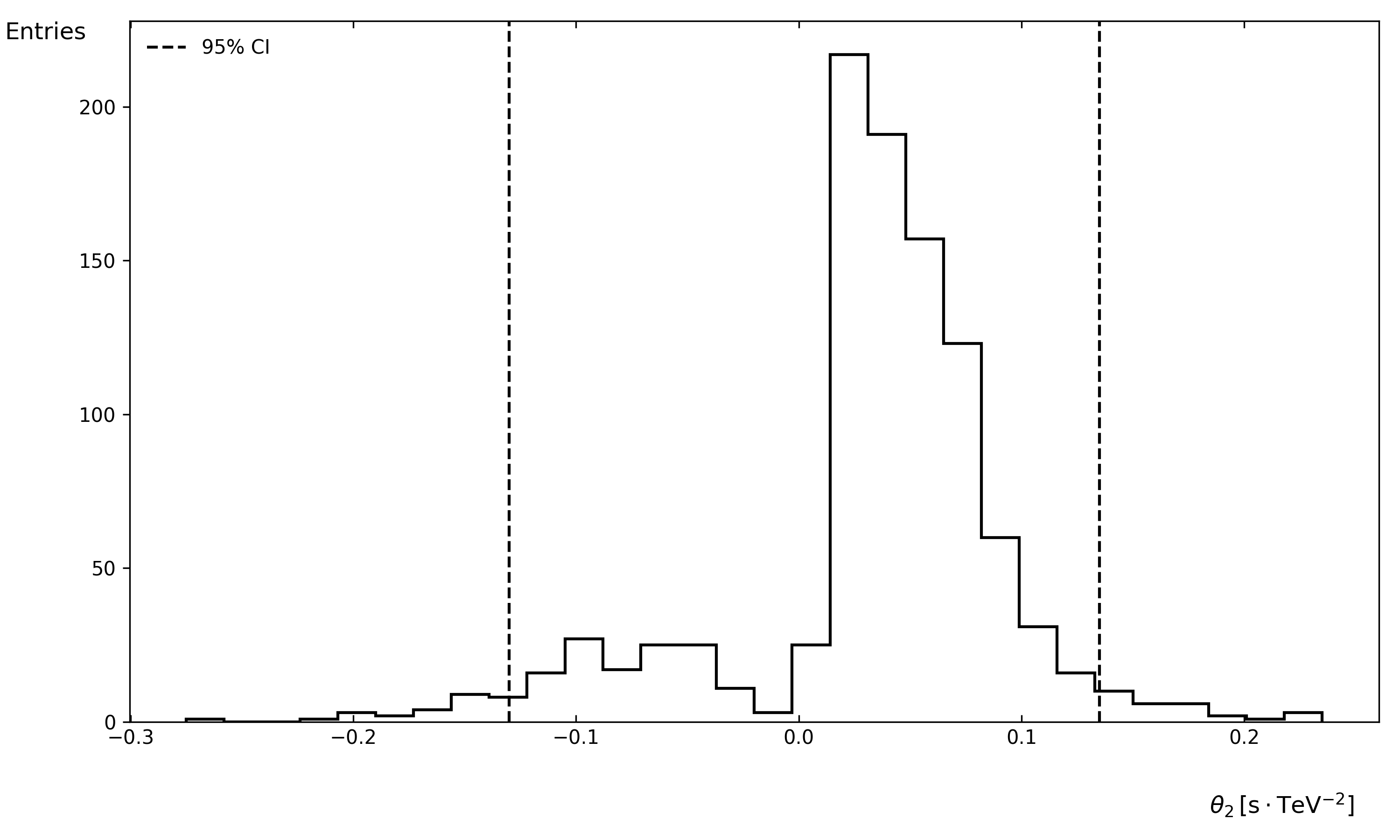}
    \caption{}
    \label{fig:bgS2}
  \end{subfigure}
  \caption{The distribution histograms of parameters $\theta_1$ and $\theta_2$ derived from the application of Knuth's optimal binning algorithm with relative Shannon entropy as the cost function}
  \label{fig:bgS}
\end{figure}

\begin{figure}[htbp]
  \centering
  \begin{subfigure}{0.45\textwidth}
    \includegraphics[width=\linewidth]{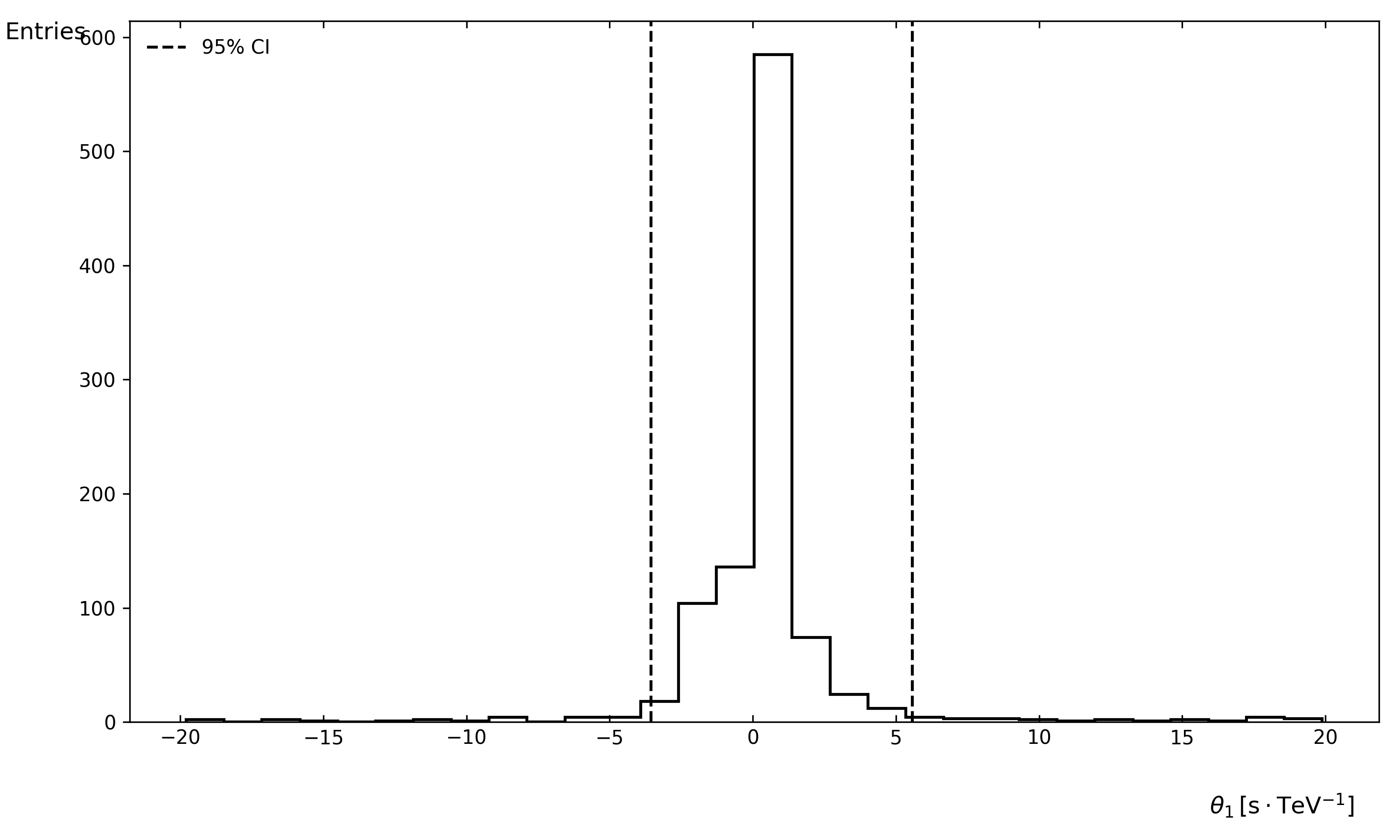}
    \caption{}
    \label{fig:bgT1_2}
  \end{subfigure}
  \hfill
  \begin{subfigure}{0.45\textwidth}
    \includegraphics[width=\linewidth]{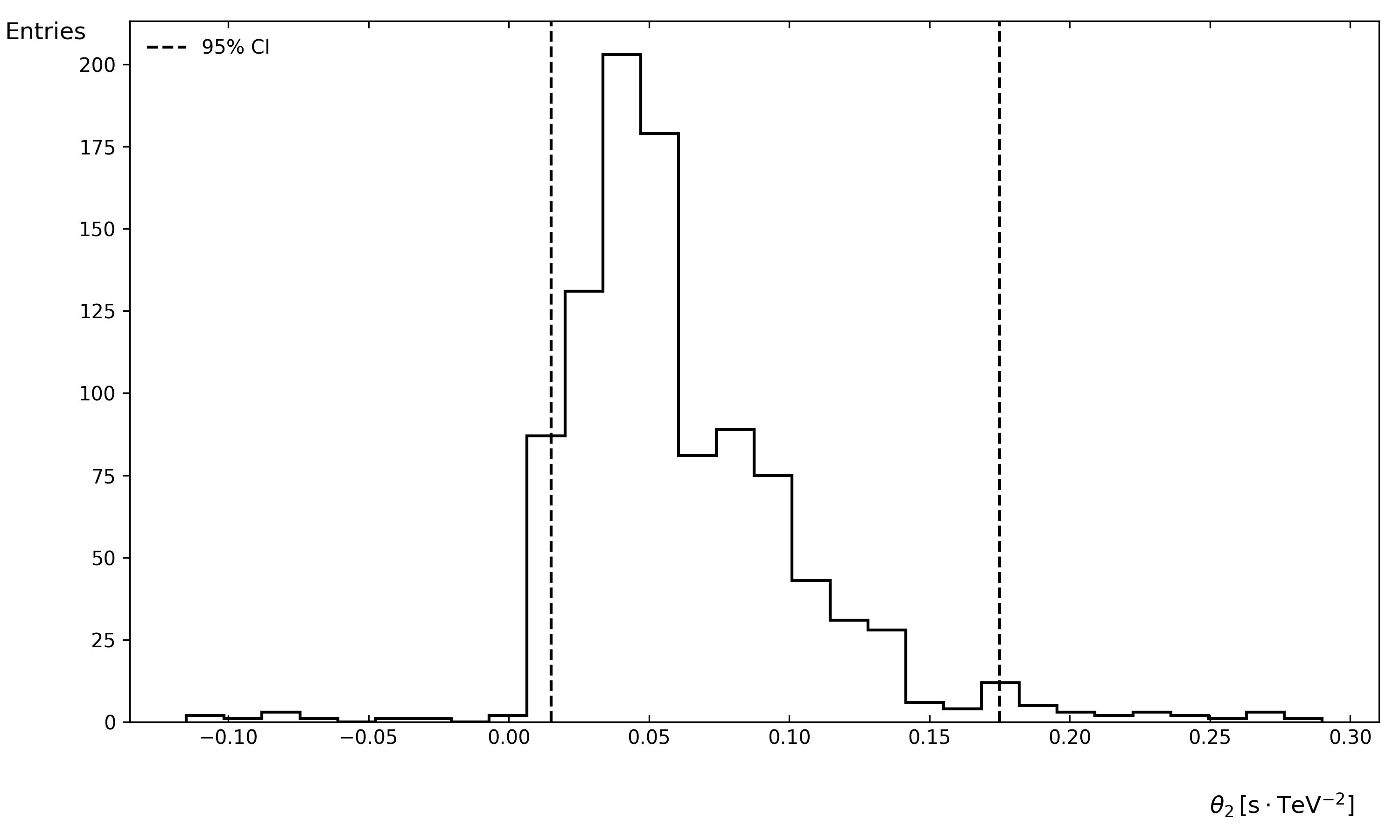}
    \caption{}
    \label{fig:bgT2_2}
  \end{subfigure}
  \caption{The distribution histograms of parameters $\theta_1$ and $\theta_2$ derived from the application of Knuth's optimal binning algorithm with relative Tsallis entropy ($q=2$) as the cost function}
  \label{fig:bgT2}
\end{figure}

\begin{figure}[htbp]
  \centering
  \begin{subfigure}{0.45\textwidth}
    \includegraphics[width=\linewidth]{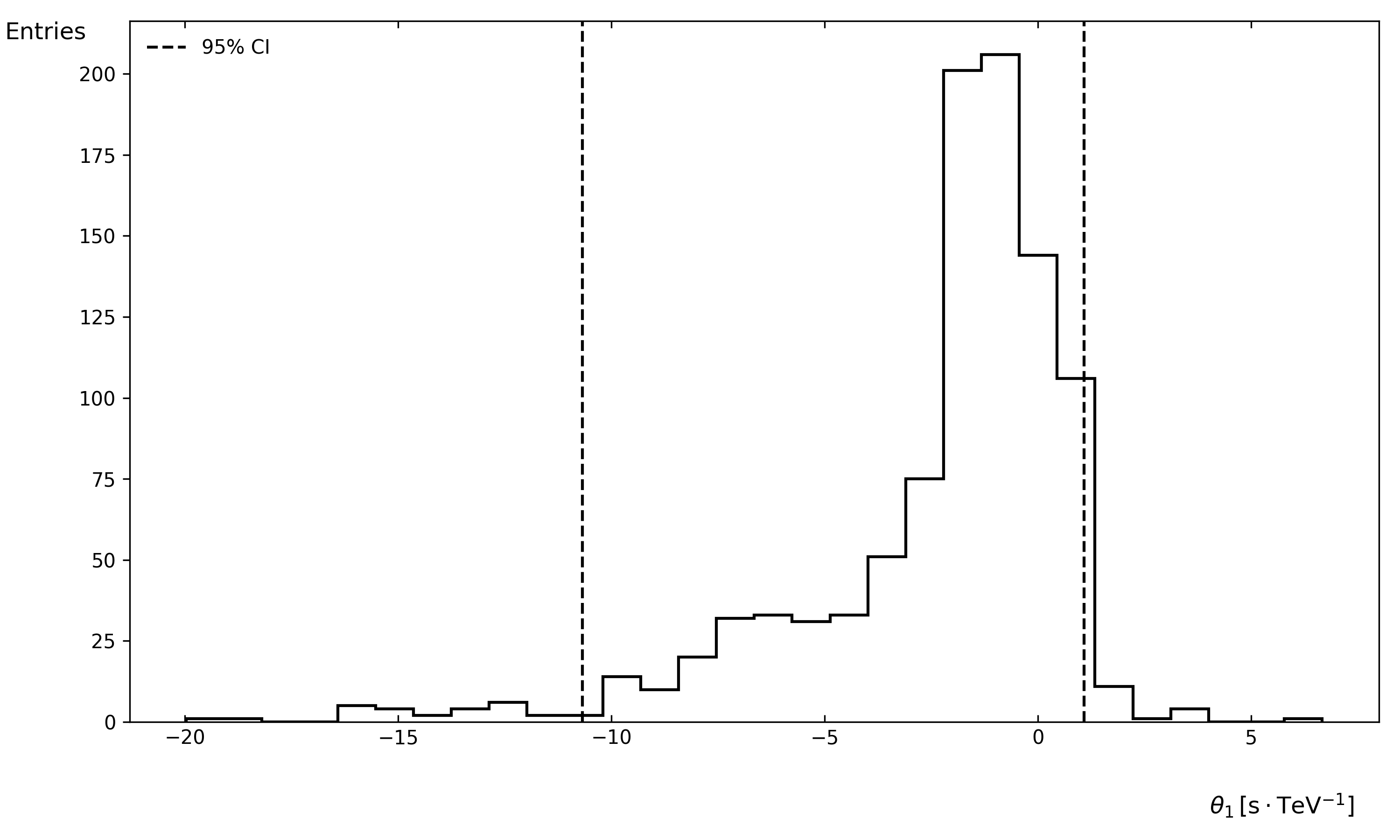}
    \caption{}
    \label{fig:bgT1_0.5}
  \end{subfigure}
  \hfill
  \begin{subfigure}{0.45\textwidth}
    \includegraphics[width=\linewidth]{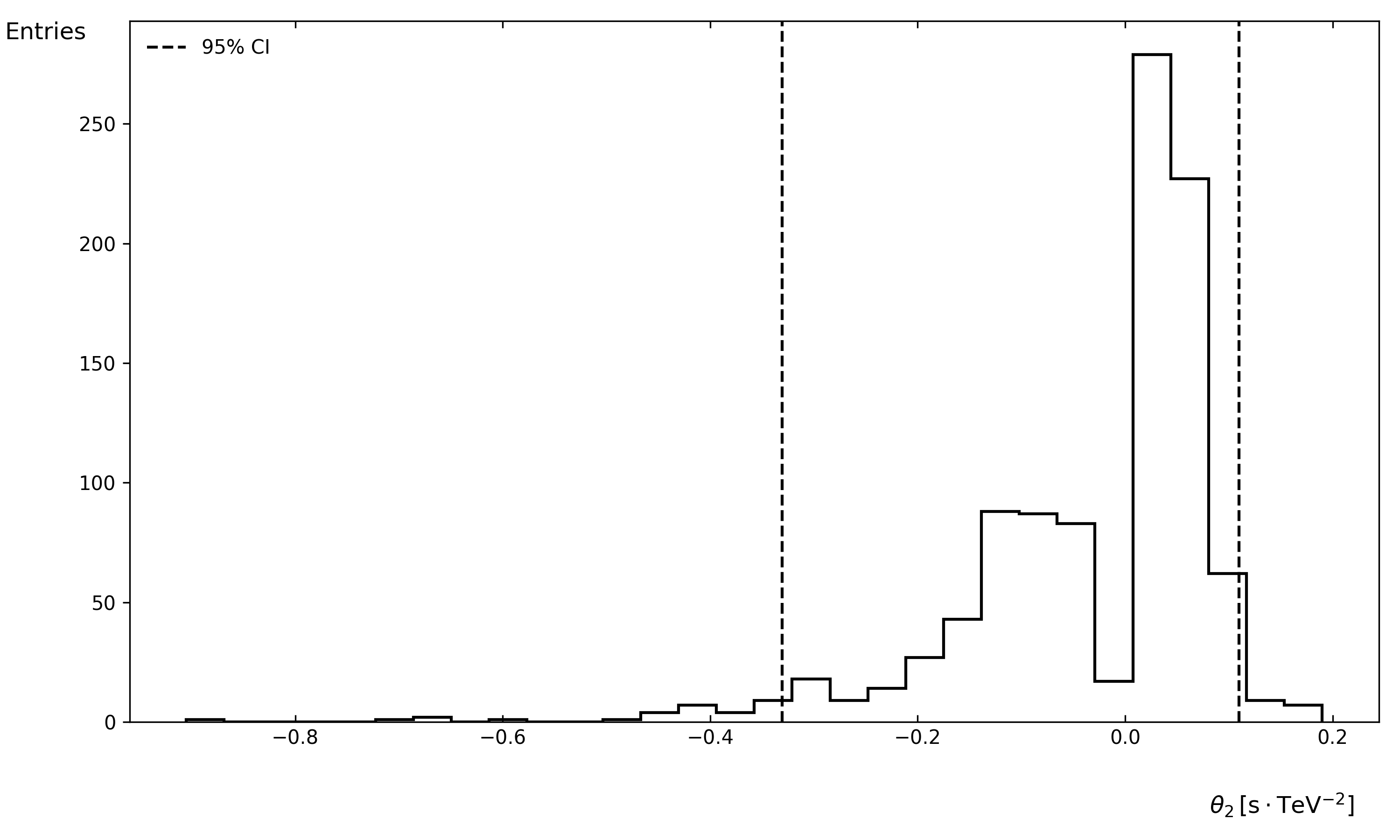}
    \caption{}
    \label{fig:bgT2_0.5}
  \end{subfigure}
  \caption{The distribution histograms of parameters $\theta_1$ and $\theta_2$ derived from the application of Knuth's optimal binning algorithm with relative Tsallis entropy ($q=0.5$) as the cost function}
  \label{fig:bgT0.5}
\end{figure}

FIGs. \ref{fig:bgS}, \ref{fig:bgT2}, and \ref{fig:bgT0.5} present the histogram of the $\theta_i$ distribution obtained after incorporating WCDA photons and applying the Knuth binning method.It can be observed that, upon employing the Knuth binning method, we directly exclude the scenario where relative R\'enyi entropy is used as the cost function. Additionally, a comparison is conducted for the parameter ($q$) in relative Tsallis entropy, revealing that the case with $q=0.5$ is superior to that with $q=2$. The physical interpretation of this finding is that our system exhibits significant non-extensivity, meaning that the total entropy of the system is not merely the sum of the entropies of its subsystems---a characteristic notably associated with R\'enyi entropy. In the analysis of gamma-ray burst (GRB) photon time series, the arrival times of photons may exhibit long-range correlations influenced by quantum gravity effects. Tsallis entropy with $q<1$ is more adept at capturing such correlations. Moreover, our adoption of relative entropy analysis reveals that the current task necessitates a focus on low-probability, high-value tail information—specifically, the KM2A photon data. This implies that the system may involve complex interactions that conventional extensive entropy measures cannot fully characterize. However, it is critical to note that the generality of this feature remains unverified. Future work should employ the DisCan method to conduct similar analyses on additional gamma-ray burst sources to validate these findings.

 Our initial motivation for introducing Tsallis entropy was its non-extensivity, which naturally describes non-local propagation effects and, when the dispersion relation is non-linear, the non-extensive framework of Tsallis entropy is more easily extended to multi-parameter optimization. This inspires us to consider modifying the dispersion relation and incorporating more quantum gravity spacetime characteristics into the model in future research.

\begin{table}[htbp]
\centering
\caption{95\% CIs for \(\theta_1\) and \(\theta_2\)}
\begin{tabularx}{0.9\textwidth}{>{\raggedright\arraybackslash}X
                               >{\centering\arraybackslash}X
                               >{\centering\arraybackslash}X
                               >{\centering\arraybackslash}X
                               >{\centering\arraybackslash}X}
\hline\hline
& \multicolumn{2}{c}{$\theta_1$ (s/TeV)} & \multicolumn{2}{c}{$\theta_2$ (s/TeV$^2$)} \\
& \scriptsize Lower & \scriptsize Upper & \scriptsize Lower & \scriptsize Upper \\
\hline
$H_{\mathrm{S}}$ & -2.545 & 1.290 & -0.380 & 0.090 \\
$H_{\mathrm{R}}(\alpha=2)$ & -5.435 & 2.380 & -0.380 & 0.240 \\
$H_{\mathrm{T}}(q=2)$ & -6.007 & 2.770 & -0.450 & 0.270 \\
\hline
\end{tabularx}
\label{tab:1}
\end{table}

\begin{table}[htbp]
\centering
\caption{95\% CIs for \(\varepsilon_{QG,1}\) and \(\varepsilon_{QG,2}\)}
\begin{tabularx}{0.9\textwidth}{>{\raggedright\arraybackslash}X
                               >{\centering\arraybackslash}X
                               >{\centering\arraybackslash}X
                               >{\centering\arraybackslash}X
                               >{\centering\arraybackslash}X}
\hline\hline
& \multicolumn{2}{c}{$\varepsilon_{QG,1}$ } & \multicolumn{2}{c}{$\varepsilon_{QG,2}$ (10$^{-7}$) } \\
& \scriptsize $\sigma=+1$ & \scriptsize $\sigma=-1$ & \scriptsize $\sigma=+1$ & \scriptsize $\sigma=-1$ \\
\hline
$H_{\mathrm{S}}$ & 2.2 & 4.4 & 2.0 & 8.3 \\
$H_{\mathrm{R}}(\alpha=2)$ & 1.0 & 2.4 & 2.0 & 3.1 \\
$H_{\mathrm{T}}(q=2)$ & 0.95 & 2.1 & 1.7 & 2.8 \\
\hline
\end{tabularx}
\label{tab:2}
\end{table}

\begin{table}[htbp]
\centering
\caption{95\% CIs for $E_{\text{QG},1}$\footnote{In some cases, the inferred linear LIV scale $E_{\text{QG},1}$ exceeds the Planck energy $E_{Pl} \approx 1.22 \times 10^{19} \text{GeV}$ (similar results also have been observed in previous references \cite{yang2024constraints,cao2024stringent}). Theoretically, this potentially implies the absence of first-order Lorentz symmetry breaking. However, there are some quantum gravity theories which have energy scales beyond $E_{Pl}$ \cite{gross1988string,amati1987superstring}. In that case, this gives a stringent constraint on the energy scale of the potential first-order Lorentz symmetry breaking. Phenomenologically, these apparent super-Planckian values possibly indicate unmitigated systematic effects. Future multi-messenger observations (such as gamma-ray bursts (GRBs) and active galactic nuclei (AGNs)) could resolve these ambiguities by reducing statistical uncertainties and constraining source-related systematics. } and $E_{\text{QG},2}$}
\begin{tabularx}{0.9\textwidth}{>{\raggedright\arraybackslash}X
                               >{\centering\arraybackslash}X
                               >{\centering\arraybackslash}X
                               >{\centering\arraybackslash}X
                               >{\centering\arraybackslash}X}
\hline\hline
& \multicolumn{2}{c}{$E_{\text{QG},1}$ (10$^{19}$ GeV) } & \multicolumn{2}{c}{$E_{QG,2}$ (10$^{12}$ GeV) } \\
& \scriptsize $\sigma=+1$ & \scriptsize $\sigma=-1$ & \scriptsize $\sigma=+1$ & \scriptsize $\sigma=-1$ \\
\hline
$H_{\mathrm{S}}$ & 2.7 & 5.4 & 2.4 & 10.0 \\
$H_{\mathrm{R}}(\alpha=2)$ & 1.3 & 2.9 & 2.4 & 3.8 \\
$H_{\mathrm{T}}(q=2)$ & 1.2 & 2.5 & 2.0 & 3.4 \\
\hline
\end{tabularx}
\label{tab:3}
\end{table}

Tables \ref{tab:1}, \ref{tab:2}, and \ref{tab:3} list the 95\% confidence intervals for the parameters \( \theta_i \), \( \varepsilon_{\mathrm{QG},n} \), and \( E_{\mathrm{QG},n} \), where \( \sigma = +1 \) represents the superluminal case and \( \sigma = -1 \) represents the subluminal case. It can be observed that for the parameter \( \theta \), regardless of whether it is first-order or second-order, and regardless of the entropy measure used as the cost function, the absolute value of the lower bound is always greater than that of the upper bound. We have obtained conclusions consistent with those of the LHAASO collaboration group \cite{yang2024constraints}. This indicates that the distribution is not symmetric, but rather biased. An additional feature observed in Figures 3–5 is that, after incorporating WCDA photons (which exhibit lower energy contrast compared to KM2A photons) and applying the Knuth binning method, the photon distribution shows a preference for subluminal propagation. This trend contrasts with the results obtained from KM2A data alone.

In Table \ref{tab:4}, it can be observed that for the case of relative Tsallis entropy with $q=2$, in the second-order scenario, both lower and upper limits for the second-order Lorentz violation energy scale are obtained. When employing the Knuth binning method and relative $H_{\mathrm{T}}(q=2)$ entropy, the following results are derived: in the superluminal case, the lower limit for $E_{\text{QG},1}$ is $2.0 \times 10^{19} \, \text{GeV}$; in the subluminal case, the lower limit for $E_{\text{QG},1}$ is $1.2 \times 10^{19} \, \text{GeV}$; additionally, in the subluminal case, the lower limit for $E_{\text{QG},2}$ is $5.2 \times 10^{12} \, \text{GeV}$ and the upper limit is $6.1 \times 10^{13} \, \text{GeV}$. These limits are determined based on the 95\% confidence interval.The results in Tables \ref{tab:4} and \ref{tab:5} demonstrate that we have further considered the influence of the afterglow of GRB 221009A and the effect of parameter adjustment in Tsallis entropy.

In the process of deriving \( E_{\mathrm{QG},n} \) from \( \varepsilon_{\mathrm{QG},n} \), we find that, regardless of whether the first-order or second-order linear delay model is used, the subluminal constraint is always stronger than the superluminal one. Specifically, the first-order energy scale is on the order of \( 10^{19} \) GeV, and the second-order energy scale is on the order of \( 10^{12} \) GeV. The constraints obtained using Shannon entropy as the cost function are stricter, indicating that it provides more comprehensive results in the absence of additional physical considerations.

\begin{table}[htbp]
\centering
\caption{ 95\% CIs for $\theta_1$ and $\theta_2$ with Knuth's binning method and relative entropy .}
\begin{tabularx}{0.9\textwidth}{>{\raggedright\arraybackslash}X
                               >{\centering\arraybackslash}X
                               >{\centering\arraybackslash}X
                               >{\centering\arraybackslash}X
                               >{\centering\arraybackslash}X}
\hline\hline
& \multicolumn{2}{c}{$\theta_1$ (s/TeV)} & \multicolumn{2}{c}{$\theta_2$ (s/TeV$^2$)} \\
& \scriptsize Lower & \scriptsize Upper & \scriptsize Lower & \scriptsize Upper \\
\hline
$H_{\mathrm{S}}$ & -7.476 & 1.586 & -0.130 & 0.135 \\
$H_{\mathrm{T}}(q=2)$ & -3.556 & 5.577 & 0.015 & 0.175 \\
$H_{\mathrm{T}}(q=0.5)$ & -10.676 & 1.080 & -0.330 & 0.110 \\
\hline
\end{tabularx}
\label{tab:4}
\end{table}

\begin{table}[htbp]
\centering
\caption{95\% CIs for $E_{\text{QG},1}$ and $E_{\text{QG},2}$ with Knuth's binning method and relative entropy .}
\begin{tabularx}{0.9\textwidth}{>{\raggedright\arraybackslash}X
                               >{\centering\arraybackslash}X
                               >{\centering\arraybackslash}X
                               >{\centering\arraybackslash}X
                               >{\centering\arraybackslash}X}
\hline\hline
& \multicolumn{2}{c}{$E_{\text{QG},1}$ (10$^{19}$ GeV) } & \multicolumn{2}{c}{$E_{QG,2}$ (10$^{12}$ GeV) } \\
& \scriptsize $\sigma=+1$ & \scriptsize $\sigma=-1$ & \scriptsize $\sigma=+1$ & \scriptsize $\sigma=-1$ \\
\hline
$H_{\mathrm{S}}$ & 0.93 & 4.4 & 7.0 & 6.8 \\
$H_{\mathrm{T}}(q=0.5)$ & 0.65 & 6.4 & 2.8 & 8.3 \\
\hline
\end{tabularx}
\label{tab:5}
\end{table}

\section{discussions and conclusions}
In this paper, we employed the DisCan method with different information entropy measures to investigate the constraints on LIV parameters using the LHAASO's KM2A observational data of GRB 221009A. Similar constraints were obtained using different entropy measures. Specifically, using the DisCan method with Shannon entropy, the 95\% confidence level (CL) lower limits for the subluminal (superluminal) scenarios were determined to be \( E_{\text{QG},1} > 5.4 (2.7) \times 10^{19} \, \text{GeV} \) for the linear case (\( n = 1 \)), and \( E_{\text{QG},2} > 10.0 (2.4) \times 10^{12} \, \text{GeV} \) for the quadratic case (\( n = 2 \)). By using the alternative Rényi entropy with \( \alpha = 2 \), the lower limits were found to be \( E_{\text{QG},1} > 2.9 (1.3) \times 10^{19} \, \text{GeV} \) and \( E_{\text{QG},2} > 3.8 (2.4) \times 10^{12} \, \text{GeV} \), while Tsallis entropy with \( q = 2 \) produced similar results to Rényi entropy. The Lorentz-violating energy scale obtained in our study is approximately of the same order of magnitude as that derived from the recent work by LHAASO \cite{cao2024stringent}. These results indicate that while the exact choice of entropy measure affects the precise values, the order of magnitude of the LIV energy scale remains consistent. As an optimization and supplement to the initial methods, we introduced WCDA photons and the Knuth binning method to enhance the original approach. This not only makes the results more diverse and engaging but also highlights the differences when various information entropies are employed as cost functions. Additionally, it underscores the further screening effect of the Knuth binning method on information entropy. Our findings indicate that R\'enyi entropy, being an additive entropy, is less suitable than Tsallis entropy for systems characterized by long-range interactions. Consequently, this insight directs our future research efforts toward a more in-depth investigation of the parameter ($q$). It should be emphasized that, at present, our findings do not definitively suggest the existence of long-range interactions in the observed gamma-ray burst spectra. To obtain more definitive signals, it is necessary to conduct similar analyses on other events, such as GRB 130427A and GRB 190114C. This aspect of the research will be undertaken in future work.

The derived constraints for the LIV parameters \( \theta_1 \) and \( \theta_2 \), as well as the corresponding energy scales \( E_{\text{QG},1} \) and \( E_{\text{QG},2} \), reveal a clear preference for subluminal scenarios, with stronger limits in the subluminal regime. Although we only used the KM2A data and lacked more detailed data from the LHAASO collaboration, this conclusion is consistent with theirs \cite{yang2024constraints}. Moreover, the energy scales corresponding to the first-order LIV (\( E_{\text{QG},1} \)) are on the order of \( 10^{19} \, \text{GeV} \), while the second-order LIV scales (\( E_{\text{QG},2} \)) are on the order of \( 10^{12} \, \text{GeV} \). The use of Shannon entropy as a cost function yields reliable
constraints, highlighting its advantage in providing more reliable results when additional physical factors are not incorporated into the model.

Based on Table 1 of \cite{wei2022tests}, the lower limits on the energy scale of LIV are initially constrained by subluminal propagation scenarios. However, as photon energies increase to the TeV range, a significant number of lower limits are derived from superluminal propagation cases. Regarding the gamma-ray burst GRB 221009A, our analysis results are consistent with the conclusions of the LHAASO collaboration \cite{yang2024constraints}. The underlying mechanism responsible for superluminal propagation, however, remains unclear, this will be the objective of our next research phase.

The LHAASO collaboration elucidates that the rapid rising and gradual decaying phases of the GRB afterglow's light curve are particularly sensitive to subluminal LIV effects \cite{yang2024constraints}. This asymmetry arises because the time-delay signatures induced by subluminal dispersion ($\sigma$ = -1) align more prominently with the steep flux variations in the early afterglow phase, thereby yielding tighter constraints compared to superluminal scenarios ($\sigma$ = +1) where dispersion effects may be diluted by the smoother temporal evolution.  Due to the rapid rise and slow decay characteristics of the afterglow, high-energy photons are predominantly concentrated in the early time periods, resulting in a higher accumulation of photons within individual bins. Our method appears to favor an even distribution of photons across each bin, thus preferring a dispersion coefficient $\theta$ that shifts the densely packed high-energy photons on the left toward the right. This likely explains why our constraints on subluminal LIV effects are stricter than those on superluminal LIV effects.

Another possible issue is that we neglect the potential effect of Extragalactic Background Light (EBL) in the present work. However, as found by the LHAASO collaboration in \cite{lhaaso2023tera}, EBL absorption in the present case exerts minimal influence on the light curves of both high-energy and low-energy photons, rendering our methodology largely insensitive to EBL effects.

It is noteworthy that our methodology is grounded in photon time series counting analysis, which differs from statistical methods that rely on parametric fitting. A promising avenue for future research lies in integrating these two approaches. Significantly, existing literature has accounted for photon emission time effects by employing data from multiple GRBs at different redshifts to conduct statistical joint analyses \cite{ellis2006robust,ellis2019robust,song2025lorentz}, thereby mitigating source-intrinsic effects. It is of great interest to combine our method to perform such multiple GRB analysis in the future, so as to effectively reduce the uncertainty of source effects.

In addition, to validate whether our statistical procedure is robust or not, it is of great significance to establish that the confidence
intervals (CIs) produced by the method reliably meet the desired coverage criteria, as illustrated in \cite{vasileiou2013constraints}. This is an another important direction of future work.


In conclusion, our study provides stringent constraints on LIV using the LHAASO observations of GRB 221009A. The results show that the subluminal scenario offers stronger constraints, and Shannon entropy provides the most reliable estimates. However, further refinement of these constraints will require consideration of additional physical factors, such as EBL absorption and intrinsic energy-dependent emission time effects, as well as the exploration of other quantum gravity models and dispersion relations in future studies.

\section*{Acknowledgements}
We extend our gratitude to Yuming Yang for useful discussions, and particularly thank Xiao-Jun Bi for his insightful discussions and valuable comments. This work is supported by the National Natural Science Foundation of China with Grant No.  12375049 and Key Program of the Natural Science Foundation of Jiangxi Province under Grant No. 20232ACB201008.  We are grateful to the anonymous referees for useful comments which greatly improved the quality of the work.

\bibliographystyle{IEEEtran}
\bibliography{ref.bib}
\end{document}